\documentstyle[psfig,bm]{nature}

\newcommand{\msun}{M$_{\odot} $}
\newcommand\etal{{\itshape{et al.}}}

\title{An increased estimate of the merger rate of double neutron 
stars from observations of a highly relativistic system}

\author{M.~Burgay\affiliation{Universit\`a degli Studi di Bologna,
Dipartimento di Astronomia, via Ranzani 1, 40127, Bologna,
Italy}, N.~D'Amico\affiliation{Universit\`a degli Studi di
Cagliari, Dipartimento di Fisica, SP Monserrato-Sestu km
0.7, 09042 Monserrato, Italy}$^{,}$\affiliation{INAF -
Osservatorio Astronomico di Cagliari, Loc. Poggio dei Pini, 
Strada 54, 09012 Capoterra, Italy}, A.~Possenti$^{3,}$\affiliation{INAF -
Osservatorio Astronomico di Bologna, via Ranzani 1, 40127, Bologna, Italy},
R.N.~Manchester\affiliation{Australia Telescope National 
Facility, CSIRO, PO Box 76, Epping, New South Waless 2121, Australia},
A.G.~Lyne\affiliation{University of Manchester, Jodrell 
Bank Observatory, Macclesfield, Cheshire, SK11 9DL, UK},
B.C.~Joshi$^{6,}$\affiliation{National Center for Radio Astrophysics, P O 
Bag 3, Ganeshkhind, Pune 411007, India}, M.A.~McLaughlin$^6$, 
M.~Kramer$^6$, J.M.~Sarkisian$^5$, F.~Camilo\affiliation{Columbia 
Astrophysics Laboratory, Columbia University, 550 West 120 th Street, 
New York, New York 10027}, V.~Kalogera\affiliation{Northwestern University, 
Department of Physics and Astronomy, Evanston, Illinois 60208,
USA}, C.~Kim$^9$, D.R.~Lorimer$^6$}
 
\mainauthor{Marta Burgay \etal}
\headertitle {Nature$|$Vol 426$|$4 December 2003$|$pp. 531\--533 }

\summary{The merger$^1$ of close binary systems containing two neutron
stars should produce a burst of gravitational waves, as predicted by
the theory of general relativity$^2$. A reliable estimate of the
double-neutron-star merger rate in the Galaxy is crucial in order to
predict whether current gravity wave detectors will be successful in
detecting such bursts. Present estimates of this rate are rather low$^{3-7}$, 
because we know of only a few double-neutron-star binaries with
merger times less than the age of the Universe. Here we report the
discovery of a 22-ms pulsar, PSR~J0737$-$3039, which is a member of a
highly relativistic double-neutron-star binary with an orbital period
of 2.4 hours. This system will merge in about 85 Myr, a time much
shorter than for any other known neutron-star binary. Together with
the relatively low radio luminosity of PSR~J0737$-$3039, this timescale
implies an order-of-magnitude increase in the predicted merger rate
for double-neutron-star systems in our Galaxy (and in the rest of the
Universe).}
\begin{document}
\maketitle

PSR~J0737$-$3039 was discovered during a pulsar search carried out using
a multibeam receiver$^8$ on the Parkes 64-m radio telescope in New South
Whales, Australia. The original detection showed a large change in apparent
pulsar period during the 4-min observation time, suggesting that
the pulsar is a member of a tight binary system. Follow-up
observations undertaken at Parkes consisting of continuous $\sim 5$-hour
observations showed that the orbit has a very short period (2.4 hrs)
and a significant eccentricity (0.088). The derived orbital parameters
implied that the system is relatively massive, probably consisting of
two neutron stars, and predicted a huge rate of periastron advance 
$\dot{\omega}$ due to effects of general relativity. Indeed, after only 
a few days of pulse-timing observations we were able to detect a 
significant value of $\dot{\omega}$. 

Interferometric observations made using the Australia Telescope Compact 
Array (ATCA) in the 20-cm 
\\
band gave an improved position and flux density 
for the pulsar. Knowledge of the pulsar position with subarcsecond precision 
allowed determination of the rotational period derivative, $\dot{P}$, and 
other parameters from the available data span. Table 1 reports 
results derived from a coherent phase fit to data taken over about five 
months. The measured value of $\dot{\omega}= 16.88^{\circ}$yr$^{-1}$ is about 
four times that of PSR~B1913$+$16 (ref. 9), previously the highest-known. If 
the observed $\dot{\omega}$ is entirely due to general relativity, it 
indicates a total system mass $M = 2.58 \pm 0.02$ \msun, where \msun~is the 
mass of the Sun.

Figure 1 shows the constraints on the masses of the pulsar and its
companion resulting from the observations so far and the mean pulse
profile as an inset. The shaded region indicates values that are ruled
out by the mass function $M_{\rm{f}}$ and the observed $\dot{\omega}$
constrains the system to lie between the two diagonal lines. Together,
these constraints imply that the pulsar mass $m_{\rm{p}}$ is less than 1.35
\msun~and that the companion mass $m_{\rm{c}}$ is greater than 1.24
\msun. The derived upper limit on $m_{\rm{p}}$ is consistent with the mass
distribution of pulsars$^{10}$ in double-neutron-star systems. The
minimum allowed companion mass implies that the companion is very
likely also to be a neutron star.

This is supported by the observed significant eccentricity of
the orbit, indicative of sudden and/or asymmetric mass loss from the
system in the explosion when the companion collapsed$^{11}$. Significant
classical contributions to $\dot{\omega}$, either from tidal 
deformation$^{12}$ or rotation-induced quadrupole moment$^{13,14}$ in the 
companion \-- effects which might be present if it were not a neutron star \--
are therefore unlikely. In fact,  $\dot{\omega}_{GR} \ll \dot{\omega}$   
would imply an unacceptably low value for the pulsar mass. Hence this system  
probably contains two neutron stars with relatively low masses in the
range 1.24 - 1.35 \msun, and the binary orbit is probably nearly edge-on 
(inclination $i \approx 90^{\circ}$). For $i = 68^{\circ}$, the implied mass 
of the pulsar is 1.24  \msun~and that of the companion is 1.35  \msun. 

A precise determination of the masses of both stars, and then
of the orbital inclination, will be obtained when the gravitational
redshift$^9$, $\gamma$, is measured. The expected minimum value (assuming an
edge-on orbit) of this parameter is 383 $\mu$s. However, its orbital
dependence is covariant with $a \sin{i}$ (Table 1) until there has 
been significant periastron precession. Whereas we currently get a formal
value of $\gamma = 400 \pm  100$ $\mu$s from a fit of the timing data, 
a useful measurement of $\gamma$ should be possible within a year.

The parameters of the PSR~J0737$-$3039 binary system, and the high timing
precision made possible by the narrow pulses of this pulsar, promise to
make this system an excellent `laboratory' for the investigation of
relativistic astrophysics, in particular of all the effects that are
enhanced by a short orbital period $P_{\rm{b}}$. Geodetic spin-precession due 
to curvature of space-time around the pulsar is expected to have a period
of 75 yr. This is four times less than the previously known
shortest spin precession period and should allow detailed mapping of
the pulsar emission beam$^{15}$ in few years. It may also have the undesirable 
effect of making the pulsar undetectable in a short time. 

The secular decay of the orbital period due to gravitational
wave emission should be measurable within about 15 months. In PSR~B1913$+$16, 
uncertainty in the orbital period derivative due to
acceleration in the Galactic potential limits the precision of this
test of the predictions of general relativity$^{16,17}$. By comparison,
PSR~J0737$-$3039 is relatively close to the Sun and the uncertainty in
Galactic acceleration will be much smaller, allowing a more precise
test. Because of the probable high orbital inclination of the system,
the delay due to the curvature of the space-time in the vicinity of
the companion (the Shapiro delay$^{18}$) should be measurable. 

The relatively small orbital eccentricity suggests that this system may
have already undergone a substantial change of the orbital elements
due to gravitational decay. Assuming that the evolution of orbital
separation and eccentricity are entirely driven by gravitational radiation 
emission$^{19}$, and adopting a value of 100 Myr for the time elapsed since 
birth$^4$, $\tau_{\rm{b}}$ (see Table 1), the orbital initially had a period 
of 3.3 h and an eccentricity of 0.119. Hence this system would already have 
experienced a $\sim 25\%$ decay of the orbital period and circularization,
which would be the largest variation of these parameters in the 
sample of the known double-neutron-star binary systems.

Of the five double-neutron-star systems known so far$^{20}$, only
three have orbits tight enough that the two neutron stars will
merge within a Hubble time. Two of them (PSR~B1913$+$16 and 
PSR~B1534$+$12) are located in the Galactic field, while the third 
(PSR~B2127$+$11C) is found on the outskirts of a globular cluster. The
contribution of globular cluster systems to the Galactic merger rate
is estimated to be negligible$^{21}$. Also, recent studies$^6$ have
demonstrated that the current estimate of the Galactic merger rate $\cal{R}$
relies mostly on PSR~B1913$+$16. So, in the following discussion, we
start by comparing the observed properties of PSR~B1913$+$16 and 
PSR~J0737$-$3039. 

PSR~J0737$-$3039 and its companion star will merge due to
the emission of gravitational waves in $\tau_{\rm{m}} \approx 85$ Myr, a 
timescale that is a factor 3.5 shorter than that for PSR~B1913$+$16 (ref. 9). 
In addition, the estimated distance for PSR~J0737$-$3039 (500 - 600 pc, based 
on the observed dispersion measure and a model for the distribution of
ionised gas in the interstellar medium$^{22}$) is an order of magnitude less 
than that of PSR~B1913$+$16. These properties have a substantial effect on 
the prediction of the rate of merging events in the Galaxy. 
\\

For a given class $k$ of binary pulsars in the Galaxy, apart from a beaming 
correction factor, the merger rate $\cal{R}$$_{\rm{k}}$ is calculated$^6$
as $\cal{R}$$_{\rm{k}} \propto N_{\rm{k}}/\tau_{\rm{k}}$. Here $\tau_{\rm{k}}$ 
is the binary pulsar lifetime defined as the sum of the time since birth$^4$, 
$\tau_{\rm{b}}$, and the remaining time before coalescence, $\tau_{\rm{m}}$,
and $N_{\rm{k}}$ is the scaling factor defined as the number of binaries in 
the Galaxy belonging to the given class. The shorter lifetime of
PSR~J0737$-$3039 ($\tau_{1913}/\tau_{0737} = (365 {\rm{~Myr}})/(185 
{\rm{~Myr}}) \approx 2$, where the subscript numbers refer to the pulsars), 
implies a doubling of the ratio $\cal{R}$$_{0737}/\cal{R}$$_{1913}$. 
A much more substantial increase results from the computation of the ratio of 
the scaling factors $N_{0737}/N_{1913}$. The luminosity $L_{400} \approx 30$ 
mJy kpc$^2$ of PSR~J0737$-$3039 is much lower than that of PSR~B1913$+$16 
($\sim 200$ mJy kpc$^2$). For a planar homogeneous distribution of pulsars in 
the Galaxy, the ratio $N_{0737}/N_{1913} \propto L_{1913}/L_{0737} \approx 6$. 
Hence we obtain $\cal{R}$$_{0737}/\cal{R}$$_{1913} \approx 12$. Including the 
moderate contribution of the longer-lived PSR~B1534$+$12 system to the total 
rate$^{4-7}$, we obtain an increase factor for the total merger rate 
$(\cal{R}$$_{0737}+\cal{R}$$_{1913}+\cal{R}$$_{1534})/(\cal{R}$$_{1913}+
\cal{R}$$_{1534})$ of about an order of magnitude. A better estimate of this 
factor and its uncertainty can be obtained using a bayesian statistical 
approach and accounting for the full luminosity function of pulsars in the 
Galaxy$^6$. 

Figure 2 dispalys an estimate of the probability density function
for the merger rate increase factor, showing a peak value of $\sim 8$ and an 
upper limit of $\sim 30$ at a $95\%$ confidence level. This result, derived in 
the simple assumption of a fixed pulsar luminosity and a uniform disk
distribution of pulsar binaries, can be refined by including the
parameters of the present survey into a simulation program modelling
survey selection effects (and hence detection probability) and the
Galactic population of pulsars. The Parkes high-latitude survey covers
the region enclosed by Galactic latitude $|b| < 60^{\circ}$ and Galactic
longitude $220^{\circ} < l < 260^{\circ}$ for a total of 6456 pointings, 
each of duration 265 s, using a bandwidth of 288 MHz and a sampling time of
125 $\mu$s. PSR~J0737$-$3039 was detected with signal-to-noise ratio of 18.7
in a standard Fourier search$^{23}$. Variation in the apparent pulsar
period due to binary motion during the discovery observation resulted
in a $30\%$ reduction in the observed signal-to-noise ratio. 

Extensive simulations (V.K. \etal, manuscrtpt in preparation) produce results 
consistent with that derived in Fig. fig2, and show that the peak of
the merger rate increase factor resulting from the discovery of 
PSR~J0737$-$3039 lies in the range 6 to 7 and is largely
independent of the adopted pulsar population model. On the other hand,
the actual predicted value of the Galactic merger rate and hence the
detection rate by gravity wave detectors depends on the shape of
the pulsar luminosity function. For the most favourable
distribution model available (model 15 of ref. 6), the updated cosmic 
detection rate for first-generation gravity wave detectors such as 
VIRGO$^{24}$, LIGO$^{25}$ and GEO$^{26}$ can be as high as 1 every 
1-2 years at $95\%$ confidence level. 

After a few years of operation of the gravity wave detectors, it
should be possible to test these predictions directly and thus place
better constraints on the cosmic population of double-neutron-star
binaries.

\vspace{3mm}
\noindent
{\footnotesize{Received 12 August; accepted 15 October 2003.\hfill}
\vspace{-1mm}
}
\rule{81mm}{0.3mm}
{\footnotesize{
\begin{itemize}
\item[ 1.] Misner, C., Thorne, K.S. \& Wheeler, J.A. in Gravitation,Chapter 36
    W.H. Freeman, New York (1973) 
\item[ 2.] Schutz, B. GW, Sources, and Physics Overview. Proceedings of the 5th
   Edoardo Amaldi Conference on Gravitational Waves. Classical and Quantum 
   Gravity, special issue (in the press)
\item[ 3.] Curran, S. J. \& Lorimer, D. R. Pulsar Statistics - Part Three - 
   Neutron Star Binaries. Mon. Not. R. Astron. Soc. 276, 347-352 (1995) 
\item[ 4.] Arzoumanian, Z., Cordes, J. M. \& Wasserman, I. Pulsar Spin 
   Evolution, Kinematics, and the Birthrate of Neutron Star Binaries. 
   Astrophys. J. 520, 696-705 (1999). 
\item[ 5.] Kalogera, V., Narayan, R., Spergel, D. N. \& Taylor, J. H. 
   The Coalescence Rate of Double Neutron Star Systems. Astrophys. J. 556, 
   340-356 (2001). 
\item[ 6.] Kim, C., Kalogera, V. \& Lorimer, D. R. The Probability Distribution
   of Binary Pulsar Coalescence Rates. I. Double Neutron Star Systems in the 
   Galactic Field. Astrophys. J. 584, 985-995 (2003). 
\item[ 7.] van den Heuvel, E.P.J. \& Lorimer, D.R. On the Galactic and and 
   cosmic merger rate of double neutron stars. Mon. Not. Royal Astron. Soc., 
   283, L37-L39 (1996) 
\item[ 8.] Staveley-Smith, L., Wilson, W. E., Bird, T. S., Disney, M. J., 
   Ekers, R. D., Freeman, K. C., Haynes, R. F., Sinclair, M. W., Vaile, R. A., 
   Webster, R. L. \& Wright, A. E. The Parkes 21 cm multibeam receiver. 
   Publications Astron. Soc. Australia, 13, 243-248 (1996). 
\item[ 9.] Taylor, J. H., Fowler, L. A. \& McCulloch, P. M. Measurements of 
   general relativistic effects in the binary pulsar PSR~1913$+$16. Nature 277,
   437-440 (1979). 
\item[10.] Thorsett, S. E. \& Chakrabarty, D. Neutron Star Mass Measurements. 
   I. Radio Pulsars. Astrophys. J. 512, 288-299 (1999).
\item[11.] Srinivasan, G. \& van den Heuvel, E.P.J. Some constraints on the 
   evolutionary history of the binary pulsar PSR~1913$+$16. Astron. \& 
   Astrophys. 108, 143-147 (1982) 
\item[12.] Roberts, D. H., Masters, A. R. \& Arnett, W. D. Determining the 
   stellar masses in the binary system containing the pulsar PSR~1913$+$16 - 
   Is the companion a helium main-sequence star?. Astrophys. J. 203, 
   196-201 (1976). 
\item[13.] Smarr, L. L. \& Blandford, R. The binary pulsar - Physical 
   processes, possible companions, and evolutionary histories. Astrophys. 
   J. 207, 574-588 (1976).
\item[14.] Wex, N. A timing formula for main-sequence star binary pulsars. 
   Mon. Not. Royal Astron. Soc. 298, 66-77 (1998) 
\item[15.] Weisberg, J.M. \& Taylor, J.H. General Relativistic Geodetic Spin 
   Precession in Binary Pulsar B1913$+$16: Mapping the Emission Beam in Two 
   Dimensions. Astrophys. J. 576, 942-949 (2002). 
\item[16.] Damour, T. \& Taylor, J.H. On the orbital period change of the 
   binary pulsar PSR~1913$+$16. Astrophys. J., 366, 501-511 (1991) 
\item[17.] Taylor, J.H. Binary pulsars and relativistic gravity. Rev. Modern 
   Phys., 66 711-719 (1994) 
\item[18.] Shapiro, I. I. Fourth Test of General Relativity. Phys. Review 
   Lett., 13, 789-791 (1964) 
\item[19.] Peters, P.C. \& Mathews, J. Gravitational radiation from point 
   masses in a keplerian orbit. Phys. Rev. 131, 435-440 (1963) 
\item[20.] Taylor, J. H. Pulsar timing and relativistic gravity. Phil. Trans. 
   Royal Soc. of London, 341, 117-134 (1992). 
\item[21.] Phinney, E. S. The rate of neutron star binary mergers in the 
   universe - Minimal predictions for gravity wave detectors. Astrophys. J. 
   380, L17-L21 (1991). 
\item[22.] Taylor, J. H. \& Cordes, J. M. Pulsar distances and the galactic 
   distribution of free electrons. Astrophys. J. 411, 674-684 (1993) 
\item[23.] Manchester, R. N., Lyne, A. G., Camilo, F., Bell, J. F., Kaspi, V. 
   M., D'Amico, N., McKay, N. P. F., Crawford, F., Stairs, I. H., Possenti, 
   A., Kramer, M. \& Sheppard, D. C. The Parkes multi-beam pulsar survey - 
   I. Observing and data analysis systems, discovery and timing of 100 
   pulsars. Mon. Not. Royal Astron. Soc. 328, 17-35 (2001)
\item[24.] Bradaschia, C. et al. in Gravitational Astronomy: Instrument Design 
   and Astrophysical Prospects (eds McClelland D.E. \& Bachor, H.A.) 110-115 
   (Elizabeth and Frederick White Research Conference Proceedings, World 
   Scientific Publishing, Singapore, 1991). 
\item[25.] Abramovici, A., Althouse, W. E., Drever, R. W. P., Gursel, Y., 
   Kawamura, S., Raab, F. J., Shoemaker, D., Sievers, L., Spero, R. E., 
   Thorne, K. S. LIGO - The Laser Interferometer Gravitational-Wave
   Observatory. Science 256, 325-333 (1992). 
\item[26.] Danzmann, K. et al. GEO 6 - a 600m laser interferometric 
   gravitational wave antenna. in First Edoardo Amaldi Conference on 
   Gravitational Wave Experiments, Eds E. Coccia, G. Pizzella \& F. Ronga 
  (World Scientific, Singapore) 100-111 (1995) 
\item[27.] Damour, T. \& Deruelle N. General relativistic celestial mechanics 
   of binary systems. II The post-Newtonian timing formula. Ann. Inst. H. 
   Poincar\'e (Phisique Th\'eorique) 44, 263-292 (1986). 
\item[28.] Manchester, R.N. \& Taylor, J.H. in Pulsars, Chapter 9. W. H. 
   Freeman, San Francisco (1977).
\end{itemize}
}}

\vspace{5mm}
\noindent
{\bf{Acknowledgements}} 
{\small{
{We thank J. Reynolds of the Parkes Observatory, and B. Sault of the ATCA, 
for prompt allocations of observing time. The Parkes
Observatory and the ATCA are part of the Australia Telescope, which is
funded by the Commonwealth of Australia for operation as a National
Facility managed by CSIRO. M.B., N.D'A. and A.P. acknowledge financial
support from the Italian Ministry of University and Research (MIUR)
under the national program `Cofin 2001'. V.K. acknowledges partial support
by a David and Lucile Packard Science \& Engineering Fellowship and a
NSF Gravitational Physics grant. D.R.L. is a University
Research fellow funded by the Royal Society.}
}}

\vspace{5mm}
\noindent
{\bf{Competing interest statement}}
{\small{The authors declare that they have no competing financial 
interests.
}}

\vspace{5mm}
\noindent
{\bf{Correspondence}} 
{\small{and requests for materials should be addressed to N.D'A. 
(damico@ca.astro.it).
}}

\begin{figure}[htbp]
\centerline{\psfig{figure=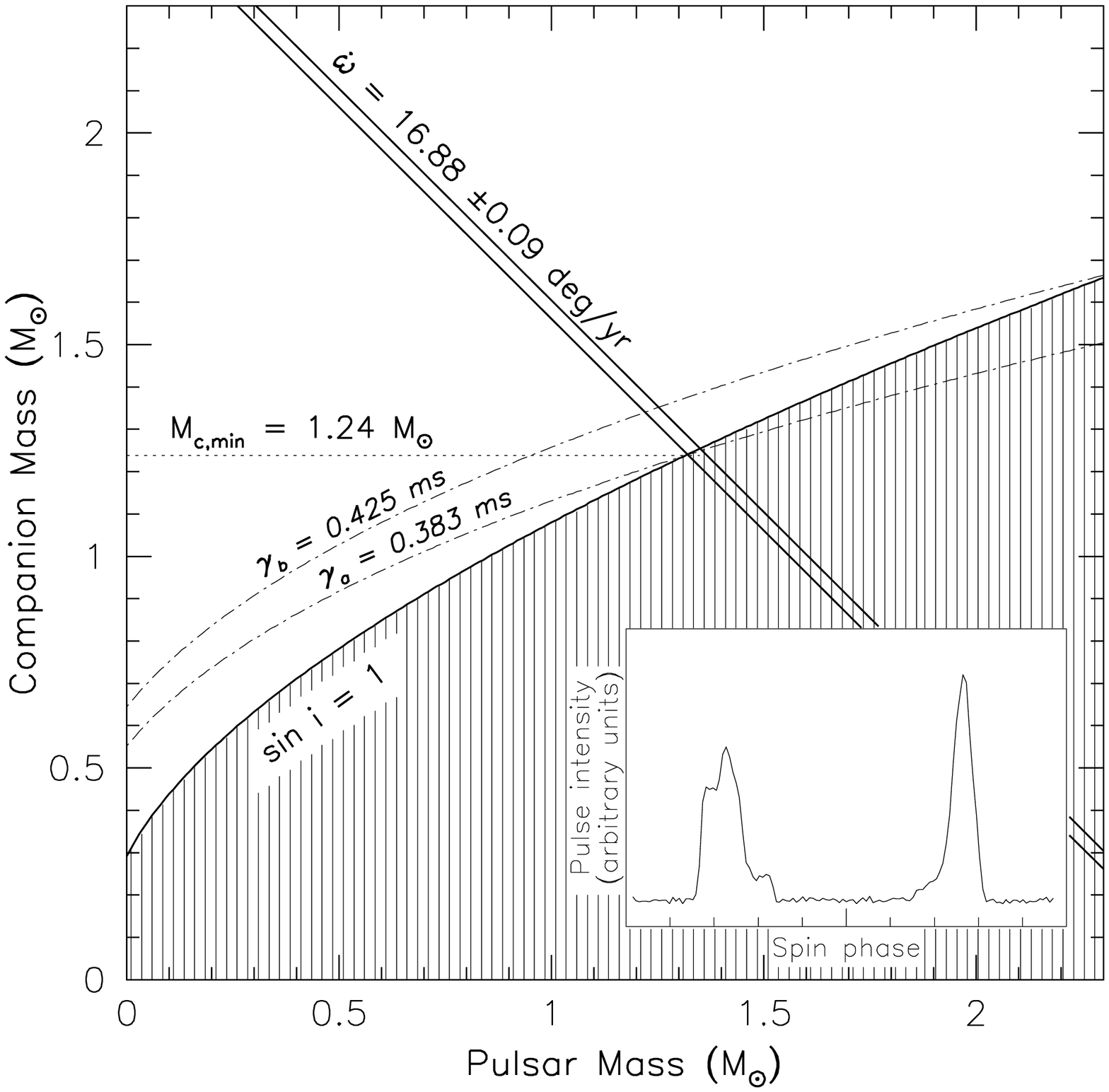,width=80mm}}
\label{fig1}
{\footnotesize{\sffamily{\bf{Figure 1}} 
Constraints on the masses of the neutron star and the companion. 
The keplerian orbital parameters give the mass function:
\begin{displaymath}
M_f=\frac{\left(m_c \sin{i}\right)^3}{\left(m_c+m_p\right)^2}
=\frac{4\pi^2\left(a \sin{i}\right)^3}{GP_b^2}
=0.291054(8) {\rm{M}}_{\odot}
\end{displaymath}
where $m_{\rm{p}}$ and $m_{\rm{c}}$ are the masses of the pulsar and
companion, $P_{\rm{b}}$ is the orbital period, and $a \sin{i}$ is the
projected semi-major axis of the pulsar orbit; the estimated
uncertainty in the last quoted digit is given in parentheses. This
provides a constraint on the minimum $m_{\rm{c}}$ as a function of
$m_{\rm{p}}$ by setting $i = 90^{\circ}$, where $i$ is the inclination
of the orbit normal to the line of sight. In addition, assuming that
the periastron advance rate $\dot{\omega}$ is entirely due to general
relativity, (that is, $\dot{\omega} =\dot{\omega}_{\rm{GR}}$), its value
can be combined with the other orbital parameters to determine the
total system mass $M = m_{\rm{c}} + m_{\rm{p}}$ using the
equation$^{27}$:
\begin{displaymath}
m_c+m_p=\frac{P_b^{5/2}}{2\pi G}
\left[\frac{\left(1-e^2\right)c^2\dot{\omega}}{6\pi}\right]^{3/2}
=2.58(2) {\rm{M}}_{\odot}
\end{displaymath}
where $e$ is the orbital eccentricity. In the $m_{\rm{c}} - m_{\rm{p}}$ 
plane, the shaded region indicates the mass values ruled
out by the mass function $M_{\rm{f}}$, and the intersection of the
boundary of this region with the $\dot{\omega}$ limits (defining a
diagonal strip in the plot) provides a further constraint on the
masses. The dash-dotted lines represent the mass values corresponding
to the relativistic parameter $\gamma$ for an edge-on orbit
($\gamma_{\rm{a}}$) and for an orbital inclination $i = 68^{\circ}$
($\gamma_{\rm{b}}$). The inset shows the mean pulse profile at 1400
MHz from a 5-h observation. There are two pulse components
separated by approximately half the 22-ms period.

}}
\end{figure}

\begin{figure}[htpb]
\vskip -2.8 truecm
\centerline{\psfig{figure=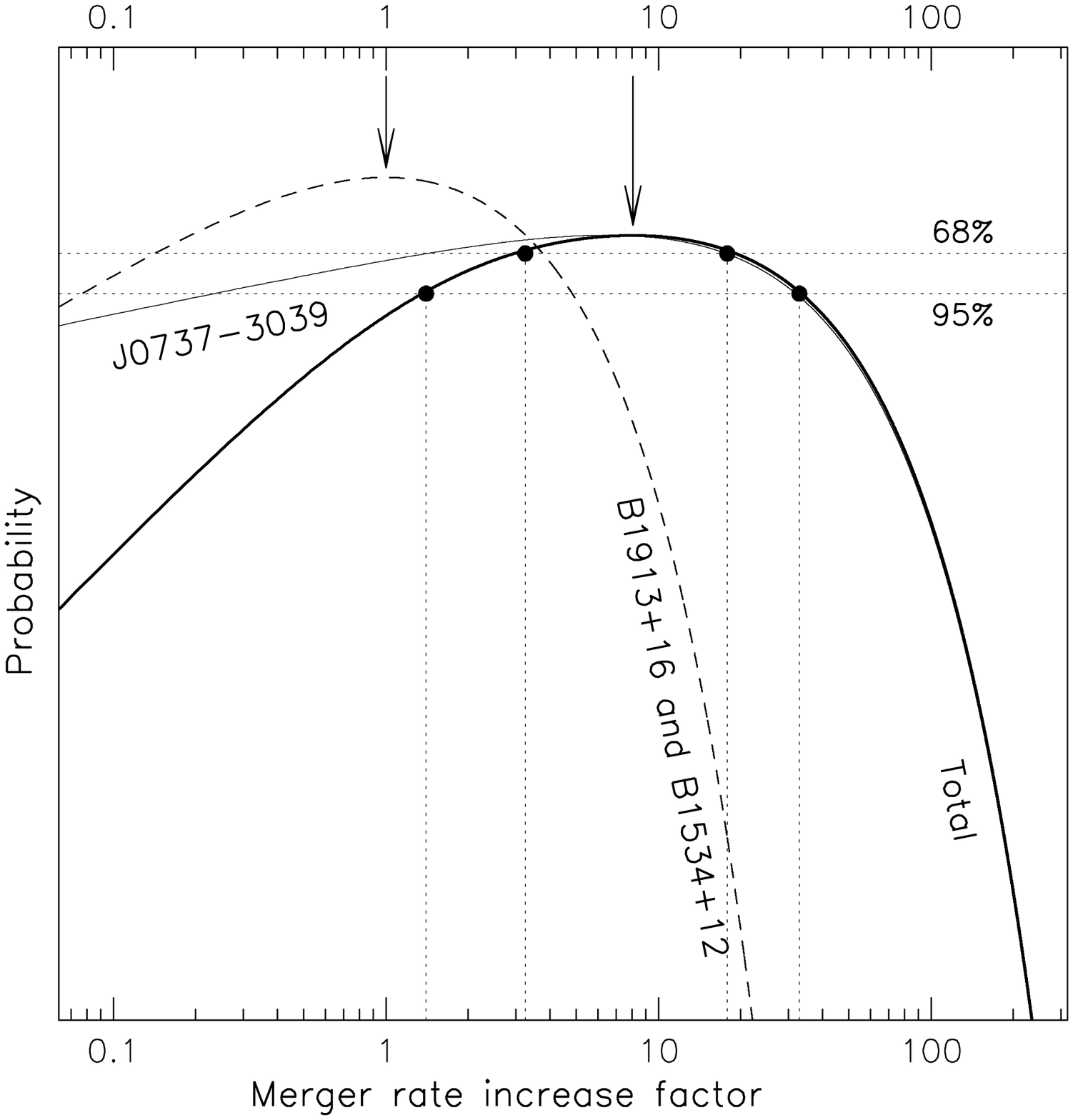,width=80mm}}
\label{fig2}
{\footnotesize{\sffamily{\bf{Figure 2}} Probability density function 
for the increase in the double-neutron-star merger rate
$(\cal{R}$$_{0737}+\cal{R}$$_{1913}+\cal{R}$$_{1534})/(\cal{R}$$_{1913}
+\cal{R}$$_{1534})$ resulting from the discovery of
PSR~J0737$-$3039. For a given class $k$ of binary pulsars in the
Galaxy, the probability density function $P(\cal{R}$$_{\rm{k}})$ for
the corresponding merger rate $\cal{R}$$_{\rm{k}}$ is obtained from
the relation$^6$ $P(\cal{R}$$_{\rm{k}})=A^2 \cal{R}$$_{\rm{k}}
e^{-A{\cal{R}}_{\rm{k}}}$ where $A=\tau_{\rm{k}}/(N_{\rm{k}}
f_{\rm{k}})$, $N_{\rm{k}}$ is a population scaling factor (see text),
$\tau_{\rm{k}}$ is the binary pulsar lifetime and $f_{\rm{k}}$ is the
correction factor due to the beamed nature of the radio pulsar
emission. The dashed line represents the reference probability
density function corresponding to the merger rate
${\cal{R}}_{{\rm{old}}}={\cal{R}}_{1913}+{\cal{R}}_{1534}$ due to
PSR~B1913$+$16 and PSR~B1534$+$12 only$^6$.  The heavy solid line
represents the probability density function corresponding to the new
merger rate
$\cal{R}$$_{\rm{new}}=\cal{R}$$_{0737}+\cal{R}$$_{1913}+\cal{R}$$_{1534}$. The
parameters adopted for PSR~J0737$-$3039 are those derived in the text
($N_{0737} =6 N_{1913}$, $\tau_{0737} =185$ Myr) and the beaming
factor for PSR~J0737$-$3039, $f_{0737}$, is chosen as the average of
the beaming factor of the other two known merging binaries
($f_{0737}=1.06 f_{1913}$). The dotted vertical lines represent the
$68\%$ and $95\%$ confidence level boundaries on the determination of
the increase factor. The dominant role of PSR~J0737$-$3039 in shaping
the new statistics of the double-neutron-star merger rate is
evident.

}}
\end{figure}

\begin{table}[htpb]
\footnotesize
\sffamily
\begin{center}
\begin{tabular}{ll}
\hline
Right Ascension (J2000) & $07^h37^m51^s.28(2)$ \\
Declination (J2000)     & $-30^{\circ}39^m40^s.3(4)$ \\
Galactic longitude	& $245^{\circ}.24$ \\
Galactic latitude	& $-4^{\circ}.50$ \\
Dispersion Measure (pc cm${-3}$) & 48.9(2) \\
Flux density at 1400 MHz (mJy) & 6.9(6) \\
Flux density at 430 MHz (mJy)  & 100(20) \\
Period, P (ms)           & 22.69937854062(6) \\
Period derivative, $\dot{P}$ (s/s) & $2.3(6) \times 10^{18}$ \\
Epoch (MJD)             & 52800.0 \\
Orbital period, $P_{\rm{b}}$ (days)  & 0.102251561(8) \\
Projected semi major axis, $a \sin{i}$ (lt-s) & 1.415176(5) \\
Eccentricity            & 0.087793(8) \\
Longitude of periastron $\omega$ & $68^{\circ}.743(9)$ \\
Epoch of periastron, $T0$ (MJD)  & 52760.807391(3) \\ 
Advance of periastron $\dot{\omega}$ ($^{\circ}$yr$^{-1}$ ) & 16.88(10) \\ 
Post-fit residual ($\mu$s) & 33 \\
Mass function, $M_{\rm{f}}$ (\msun) & 0.291054(8) \\
Total system mass, $M$ (\msun) & 2.58(2) \\
Characteristic age, $\tau_{\rm{c}}$ (Myr) & 160(50) \\
Age since birth $\tau_{\rm{b}}$ (Myr) & 100(50) \\
Surface dipole magnetic field strength, $B$ (G) & $7.3(9) \times 10^9$ \\
Expected gravitational redshift parameter, $\gamma$ ($\mu$s) & $> 383$ \\
Expected orbital period derivative (s s$^{-1}$) & $-1.24 \times 10^{-12}$ \\
\hline
\end{tabular}
\label{tab1}
\vspace{5mm}
\parbox{88mm}{\footnotesize{\sffamily Table 1 {\bf{Measured and derived 
parameters for PSR J0737 3039}}. The pulsar
position is obtained from observations made with the ATCA used in pulsar-gating
mode. Uncertainties are given in parentheses and refer to the last
quoted digit. The flux at 1400 MHz is derived from the ATCA
observations, while the flux at 430 MHz is obtained from Parkes
observations with the flux value being derived from the observed
signal-to-noise ratio and the estimated receiver sensitivity. Best-fit
timing parameters are derived from arrival-time data by minimising the
$\chi^2$ function using the program {\ttfamily{TEMPO}} (see
{\ttfamily{http://pulsar.princeton.edu/tempo}}) with the 
relativistic binary model of ref. 27. For all the timing parameters except 
$P_{\rm{b}}$, $\dot{\omega}$ and $\dot{P}$, uncertainties are twice the formal-fit 
1$\sigma$ error. Because of the short time span, $\dot{\omega}$ is strongly 
covariant with other orbital parameters and the nominal 1$\sigma$ error 
resulting from the phase coherent fit of the pulse time-of-arrival might 
not represent the actual 1$\sigma$ statistical uncertainty. By producing 
contours of constant $\chi^2$ values as a function of $P_{\rm{b}}$ and  
$\dot{\omega}$, we have verified that an uncertainty of
the order of four times the nominal 1$\sigma$ error represents the
uncertainty range for these parameters. The uncertainty in $\dot{P}$ is due
to the residual covariance with the position terms resulting from the
available uncertainty in the celestial coordinates obtained at the
ATCA. The characteristic age $\tau_{\rm{c}}$ and the surface dipole magnetic field
strength $B$ are derived from the observed $P$ and $\dot{P}$ using the 
relations$^{28}$ $\tau_{\rm{c}}=0.5 P/\dot{P}$ and $B=3.2\times 10^{19}
(P\dot{P})^{1/2}$. The age since birth, $\tau_{\rm{b}}$, is the time 
since the binary pulsar left the spin-up line$^4$.}}
\end{center}
\end{table}

\end{document}